\documentclass[reprint,amsmath,amssymb,aps,pra]{revtex4-2}

\usepackage{graphicx}
\usepackage{braket}
\usepackage{tikz}
\usepackage{quantikz}
\usepackage[colorlinks=true,urlcolor=blue,citecolor=blue,linkcolor=blue]{hyperref}

\begin{document}

\title{Loss of the quadratic speedup in Grover search under local and coherent noise}

\author{Imama Tul Birrah Khan}
\author{Muhammad Faryad}
 \email{muhammad.faryad@lums.edu.pk}
\affiliation{Department of Physics, Lahore University of Management Sciences, Lahore 54792, Pakistan}

\date{\today}

\begin{abstract}
Grover's algorithm searches an unstructured database with a number of oracle queries that grows only as the square root of its size. Noise erodes this advantage, and we ask, analytically, how quickly. When a depolarizing channel acts on the whole register after every gate layer, it commutes with the gates, and the noisy state is the noiseless one with a coherent weight that decays geometrically with depth. The Grover rotation is not slowed. Its weight decays, the best stopping point moves earlier, and the expected query cost keeps its square-root scaling only while the accumulated noise over the circuit stays small. For independent errors on the qubits we show that the ideal Grover operator freezes any amplitude pushed out of the two-dimensional search plane. This gives the exact effect of a single error: phase flips do most damage early in the search, bit flips late, and combined errors at any time. For large databases it also reduces the noisy search to Bloch-type equations for the part of the state that stays in the plane, in which the algorithm is a rotation and the noise appears as leakage and dephasing. We derive the rates for bit-flip, phase-flip, depolarizing, and amplitude-damping noise and solve the equations. Local noise acts, to leading order, like a global channel strengthened by the number of qubits, so the tolerable error per qubit falls faster than the inverse square root of the database size by one power of the register size. For coherent errors in the rotation angle we bound the success probability by a band that widens linearly with depth. Our results recover, in closed form, scalings found earlier for particular noise models, and they explain why some errors are more harmful than others.
\end{abstract}

\maketitle

\section{\label{sec:intro}Introduction}

Qubits are fragile. They couple to their environment, lose coherence, and the information they carry degrades. This is the central obstacle to practical quantum computation \cite{nielsen2010}. Any assessment of a quantum algorithm on real hardware therefore has to ask how much noise the algorithm can absorb before its advantage disappears.

Grover's search algorithm is a natural test case \cite{grover1996,grover1997}. It finds a marked item in an unstructured database of $N=2^{n}$ entries using about $\sqrt{N}$ queries to an oracle, whereas any classical strategy needs about $N$ queries, and no quantum algorithm can do better \cite{bennett1997,zalka1999}. The mechanism is amplitude amplification: each iteration rotates the state of the register a little further toward the marked state. The same mechanism underlies a wide range of oracle-based algorithms, from database search to combinatorial optimization \cite{nielsen2010,brassard2002,boyer1998}. The algorithm is simple, its noiseless behavior is known exactly, and its speedup depends on maintaining coherence over many iterations. It is therefore a good place to study how noise accumulates with circuit depth.

Real devices suffer from many noise processes \cite{resch2021}. They fall into two broad classes. Incoherent noise randomizes the state. Its standard model is the depolarizing channel, which with probability $p$ replaces the state by the maximally mixed state and with probability $1-p$ leaves it alone. Because the maximally mixed state is the average over all Pauli errors, the depolarizing channel captures, in an averaged sense, the bit-flip, phase-flip, and combined errors that other Pauli channels describe one at a time. Coherent noise, on the other hand, does not randomize the state. It rotates it by slightly the wrong amount. Calibration errors, control imprecision, and crosstalk make the oracle and the diffusion operator deviate from the ideal unitaries \cite{preskill2018,wallman2016}, and such errors can add up systematically from one iteration to the next.

Noisy Grover search has a long history. Early studies showed that random errors in the gates reduce the success probability without slowing the algorithm down \cite{pablonorman1999}, that systematic errors in the phase inversions limit the size of a database that can be searched reliably to about the inverse square of the error \cite{long2000}, and that random phase errors in the oracle must shrink with the database size for the search to remain scalable \cite{shenvi2003}. Azuma treated phase errors on the individual qubits perturbatively \cite{azuma2002}. For depolarizing noise, numerical studies found an exponential decay of the success probability \cite{salas2008}, and Cohn and co-workers derived tolerance thresholds for both a total and a local depolarizing channel and found the local case more damaging \cite{cohn2016}. From the side of complexity theory, a faulty oracle rules out any asymptotic quantum speedup \cite{regev2008}, quantum search keeps its advantage over classical search only for weak noise \cite{gawron2012}, and lower bounds show that a depolarizing oracle forces a query cost proportional to the noise rate times the database size \cite{vrana2014,rosmanis2023}. More recent work has examined localized dephasing \cite{reitzner2019}, time-correlated local noise \cite{mandal2023}, static systematic noise through random-matrix theory \cite{dowarah2025}, error correction \cite{botsinis2016}, and noise mitigation with quantum switches \cite{srivastava2025}.

Two things are still missing from this picture. The results are spread over different noise models, error placements, and figures of merit, which makes it hard to compare channels or to see why some errors are more harmful than others. And many of the available results are bounds or thresholds rather than explicit expressions for the success probability and the cost of the search. In this paper we give a single analytical treatment of Grover's algorithm under local incoherent noise and under coherent rotation errors, with every noise channel placed after every gate layer of the same circuit, so that gate errors and the idle decoherence between gates are both included.

We first solve exactly a reference model in which the whole register is depolarized at once. The noisy state separates into the noiseless Grover state, scaled by a weight that decays geometrically with the circuit depth, and a fully mixed part. From it we obtain in closed form the optimal number of iterations, the maximum success probability, and the expected number of oracle queries, and we show that the query cost attains the linear scaling allowed by the lower bounds of Refs.~\cite{vrana2014,rosmanis2023} once the noise is strong enough. We then show that for errors on individual qubits the geometry of the algorithm does most of the work. The Grover operator acts as a simple sign flip on everything outside the two-dimensional plane in which the search takes place, so any amplitude that an error pushes out of the plane is frozen. This gives the exact effect of a single error of each type, recovers Azuma's perturbative constant for phase flips \cite{azuma2002}, and, for a large database, reduces the noisy search to a closed set of linear equations for the part of the state that remains in the plane. We derive the rates in these equations for bit-flip, phase-flip, depolarizing, and amplitude-damping noise, solve them, and obtain a noise budget that agrees with the local-depolarizing threshold of Ref.~\cite{cohn2016} and extends it to the other channels. Finally, for coherent noise we bound the success probability under rotation errors of fixed maximum size, recover the scaling of Ref.~\cite{long2000}, and connect it with the case of random errors \cite{shenvi2003}.

The paper is organized as follows. Section~\ref{sec:noiseless} reviews the noiseless algorithm in density matrix language and fixes the notation. Section~\ref{sec:depol} sets up the noise model, solves the global depolarizing case exactly, analyzes single errors on individual qubits, derives and solves the equations for the Grover plane under local noise, and gives the resulting noise budget. Section~\ref{sec:rot} treats noisy rotations. Section~\ref{sec:concl} draws the threads together. Part of the analytical work was carried out with the assistance of an AI model, as described in the Acknowledgments.

\section{\label{sec:noiseless}Noiseless Grover's algorithm}

\begin{figure}[t]
    \centering
\begin{tikzpicture}
\node[scale=1.2]{
    \begin{quantikz}[row sep=0.3mm]
        \lstick{$\ket{0}$}  \slice{$\rho_1$} &\gate{H}\slice{$\rho_2$}
        & \gate[4,nwires=3]{U_f}\slice{$\rho_3$}
        \gategroup[4,steps=2,style={dashed,rounded corners,fill=blue!20, inner xsep=2pt},background,label style={label position=below,anchor=north,yshift=-0.2cm}]{{repeated $r_0$ times}}
        & \gate[4,nwires=3]{V = HU_{0}H}\slice{$\rho_f$}
        &  \\
        \lstick{$\ket{0}$} & \gate{H} & & &  \\
        & \vdots & \vdots & \vdots & \\
        \lstick{$\ket{0}$} & \gate{H} & & &
    \end{quantikz}
};
\end{tikzpicture}
    \caption{Noiseless Grover circuit \cite{nielsen2010}. An $n$-qubit register is initialized to $\ket{0}^{\otimes n}$, and a layer of Hadamard gates prepares the equal superposition. Each Grover iteration applies the oracle $U_f$, which flips the sign of the marked state, and the diffusion operator $V = H^{\otimes n}U_{0}H^{\otimes n}$, which reflects the amplitudes about their mean. The iteration is repeated $r_0$ times. The dashed lines mark the density matrices $\rho_1,\dots,\rho_f$ used in the text.}
    \label{grover_circuit}
\end{figure}
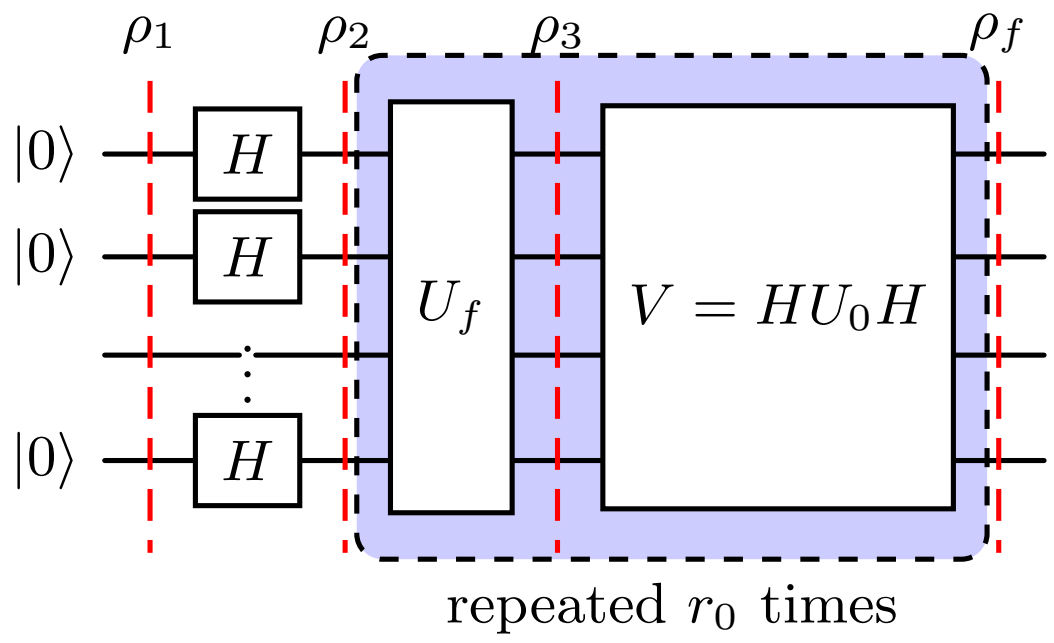

Figure~\ref{grover_circuit} shows the circuit. The register is initialized to $\ket{0}^{\otimes n}$, brought into an equal superposition by a layer of Hadamard gates, and then subjected to $r_0$ Grover iterations. Each iteration consists of the oracle $U_f$ followed by the diffusion operator $V$. We trace the density matrix through one iteration; the labels $\rho_1,\dots,\rho_f$ refer to the slices marked in the figure.

The register starts in the computational basis state
\begin{equation}
\begin{aligned}
\ket{\psi_0} &= \ket{0}^{\otimes n}, \\
\rho_1 &= \ket{0}^{\otimes n}\bra{0}^{\otimes n}.
\end{aligned}
\end{equation}
The Hadamard layer produces the equal superposition
\begin{equation}
\begin{aligned}
\ket{s} &= H^{\otimes n}\ket{0}^{\otimes n} = \frac{1}{\sqrt{N}}\sum_{x=0}^{N-1}\ket{x}, \\
\rho_2 &= \ket{s}\bra{s}.
\end{aligned}
\end{equation}
Let $\ket{w}$ be the marked state and $\ket{w^\perp}$ the normalized equal superposition of the $N-1$ unmarked states. In this basis
\begin{equation}
\begin{aligned}
\ket{s} &= \cos\alpha\,\ket{w^\perp} + \sin\alpha\,\ket{w}, \\
\sin\alpha &= \frac{1}{\sqrt{N}},
\end{aligned}
\label{eq:alpha}
\end{equation}
so that $\alpha$ is the initial angle between $\ket{s}$ and the unmarked subspace. For a large database $\alpha\approx 1/\sqrt{N}$ is small. Both the oracle and the diffusion operator map the plane spanned by $\ket{w^\perp}$ and $\ket{w}$ onto itself, so the whole algorithm takes place in this plane. This is what makes a geometric picture of the algorithm possible, and we will use it throughout.

The oracle evaluates the indicator function of the marked item,
\begin{equation}
f(x) =
\begin{cases}
1, & x = w,\\
0, & x \neq w,
\end{cases}
\end{equation}
and encodes it as a phase,
\begin{equation}
U_f \ket{x} = (-1)^{f(x)} \ket{x}.
\end{equation}
In operator form,
\begin{equation}
U_f = \sum_{x=0}^{N-1} (-1)^{f(x)} \ket{x}\bra{x}
    = \sum_{x\neq w} \ket{x}\bra{x} - \ket{w}\bra{w}.
\end{equation}
Since the projectors onto the computational basis sum to the identity, $\sum_{x} \ket{x}\bra{x} = I$, the oracle takes the compact form
\begin{equation}
U_f = I - 2\ket{w}\bra{w}.
\label{eq:oracle}
\end{equation}
It is a reflection about the hyperplane orthogonal to $\ket{w}$: only the marked state acquires a sign. In the plane of Fig.~\ref{fig:noisy_angles} below, it reflects the state about the $\ket{w^\perp}$ axis. Acting on $\rho_2$ it gives
\begin{equation}
\begin{aligned}
\rho_3 &= U_f\rho_2U_f \\
&= \cos^2\alpha\,\ket{w^\perp}\bra{w^\perp}
   + \sin^2\alpha\,\ket{w}\bra{w}\\&\quad
   - \cos\alpha \sin\alpha\,(\ket{w^\perp}\bra{w}+\ket{w}\bra{w^\perp}).
\end{aligned}
\label{eq:rho3}
\end{equation}
The oracle changes only the sign of the coherences; the populations are untouched.

The diffusion operator is built from the reflection about the all-zeros state,
\begin{equation}
U_{0} = 2\ket{0}\bra{0} - I,
\end{equation}
conjugated by Hadamard layers,
\begin{equation}
V = H^{\otimes n} U_{0} H^{\otimes n}.
\end{equation}
Since $H^{\otimes n}\ket{0}^{\otimes n} = \ket{s}$, this is
\begin{equation}
V = 2\ket{s}\bra{s} - I,
\label{eq:diffuser}
\end{equation}
a reflection about the equal superposition. It inverts every amplitude about the mean amplitude, and so amplifies the amplitude of the marked state, which was singled out by the sign flip. In the $\{\ket{w^\perp},\ket{w}\}$ basis,
\begin{equation}
\begin{aligned}
    V ={}& \cos 2\alpha \big(\ket{w^\perp}\bra{w^\perp} + \ket{w}\bra{w}\big)
    \\&+ \sin 2\alpha \big(\ket{w}\bra{w^\perp} + \ket{w^\perp}\bra{w}\big).
\end{aligned}
\end{equation}
Applying $V$ to $\rho_3$ completes one iteration,
\begin{equation}
\begin{aligned}
\rho_f &= V\rho_3V \\
&= \sin^2 3\alpha\,\ket{w}\bra{w}
 + \cos^2 3\alpha\,\ket{w^\perp}\bra{w^\perp}\\&\quad
 + \frac{\sin 6\alpha}{2}\,(\ket{w^\perp}\bra{w}+\ket{w}\bra{w^\perp}).
\end{aligned}
\label{eq:rhof_noiseless}
\end{equation}
Comparing with Eq.~\eqref{eq:alpha}, one iteration of the Grover operator $G = VU_f$ has rotated the state from angle $\alpha$ to angle $3\alpha$, that is, by $2\alpha$ toward $\ket{w}$. Two reflections make one rotation: the oracle reflects the state to the angle $-\alpha$ below the $\ket{w^\perp}$ axis, and the diffusion operator reflects it about $\ket{s}$, which sits at angle $\alpha$, up to $3\alpha$. After $r$ iterations,
\begin{equation}
\ket{\psi_r} = G^r\ket{s} = \cos[(2r+1)\alpha]\ket{w^\perp} + \sin[(2r+1)\alpha]\ket{w},
\label{eq:psir}
\end{equation}
and the probability of measuring $\ket{w}$ is
\begin{equation}
P_r^{(0)} = \sin^2[(2r+1)\alpha].
\label{eq:P0}
\end{equation}
It is maximal when $(2r+1)\alpha$ is closest to $\pi/2$, which for large $N$ gives
\begin{equation}
r_0 \approx \frac{\pi}{4}\sqrt{N} - \frac{1}{2}.
\label{eq:r0}
\end{equation}
This is the origin of the $\sqrt{N}$ query complexity. Note that for small registers the integer $r_0$ can overshoot $\pi/2$ slightly; for four qubits, for example, $r_0=3$ and $(2r_0+1)\alpha$ exceeds $\pi/2$ by about $0.2$ radians. This detail will matter in Sec.~\ref{sec:rot}.

\section{\label{sec:depol}Grover's algorithm under Pauli noise}

\subsection{Pauli channels}

The errors that a qubit suffers between two ideal gates can always be expanded in the Pauli operators. The general single-qubit Pauli channel is \cite{nielsen2010}
\begin{equation}
\begin{aligned}
    \mathcal{E}(\rho) &= (1 - p)\, \rho + p_x\, X\rho X + p_y\, Y\rho Y + p_z\, Z\rho Z , \\
    p &= p_x + p_y + p_z ,
\end{aligned}
    \label{eq:pauli}
\end{equation}
where $p$ is the total probability that an error of any kind occurs. Three special cases are standard. The bit-flip channel has $(p_x,p_y,p_z)=(p,0,0)$: with probability $p$ the qubit is flipped. The phase-flip channel has $(0,0,p)$: with probability $p$ the relative sign of $\ket{0}$ and $\ket{1}$ is reversed. The depolarizing channel has $(p/3,p/3,p/3)$: all three errors are equally likely, which is the same as replacing the qubit by the maximally mixed state $I/2$ with probability $4p/3$. We compare the three at equal total error probability $p$, so that differences between them reflect the kind of error and not its frequency.

\begin{figure*}[t]
    \centering
    \[
    \resizebox{0.85\textwidth}{!}{
    \begin{quantikz}[row sep=0.3mm]
                \lstick{$\ket{0}$} \slice{$\rho_1'$}
                    & \gate{H}\slice{$\rho_2'$}
                    & \gate[wires=1]{\mathcal{E}}\slice{$\rho_3'$}
                    & \gate[4,nwires=3]{U_f}\slice{$\rho_4'$}
                    \gategroup[4,steps=7,style={dashed,rounded corners,fill=blue!20, inner xsep=2pt},background,label style={label position=below,anchor=north,yshift=-0.2cm}]{{repeated $r$ times}}
                    & \gate[wires=1]{\mathcal{E}}\slice{$\rho_5'$}
                    & \gate{H}\slice{$\rho_6'$}
                    & \gate[wires=1]{\mathcal{E}}\slice{$\rho_7'$}
                    & \gate[4,nwires=3]{U_{0}}\slice{$\rho_8'$}
                    & \gate[wires=1]{\mathcal{E}}\slice{$\rho_9'$}
                    & \gate{H}\slice{$\rho_f'$}
                    &  \\
                \lstick{$\ket{0}$} & \gate{H} &\gate[wires=1]{\mathcal{E}} & & \gate[wires=1]{\mathcal{E}}& \gate{H}& \gate[wires=1]{\mathcal{E}}&& \gate[wires=1]{\mathcal{E}}& \gate{H}&  \\
                & \vdots & \vdots & \vdots & \vdots & \vdots & \vdots & \vdots & \vdots & \vdots & \\
                \lstick{$\ket{0}$} & \gate{H} &\gate[wires=1]{\mathcal{E}} & & \gate[wires=1]{\mathcal{E}}& \gate{H}& \gate[wires=1]{\mathcal{E}}&& \gate[wires=1]{\mathcal{E}}& \gate{H}&
            \end{quantikz}
    }
    \]
    \caption{Noisy Grover circuit. A noise channel $\mathcal{E}$ acts on every qubit after the initial Hadamard layer and after the oracle $U_f$, the first Hadamard layer, and the reflection $U_0$ of every iteration. It models both gate errors and the idle decoherence between gates. One iteration consists of the oracle and the diffusion operator $V = H^{\otimes n}U_{0}H^{\otimes n}$, and is repeated $r$ times. In Sec.~\ref{sec:global} the $n$ channels of a layer are replaced by a single global depolarizing channel on the register; in Secs.~\ref{sec:single} to \ref{sec:solutions} they are independent channels on the qubits. The dashed lines mark the density matrices $\rho_i'$ followed through the circuit in the text.}
    \label{fig:noisy grover}
\end{figure*}
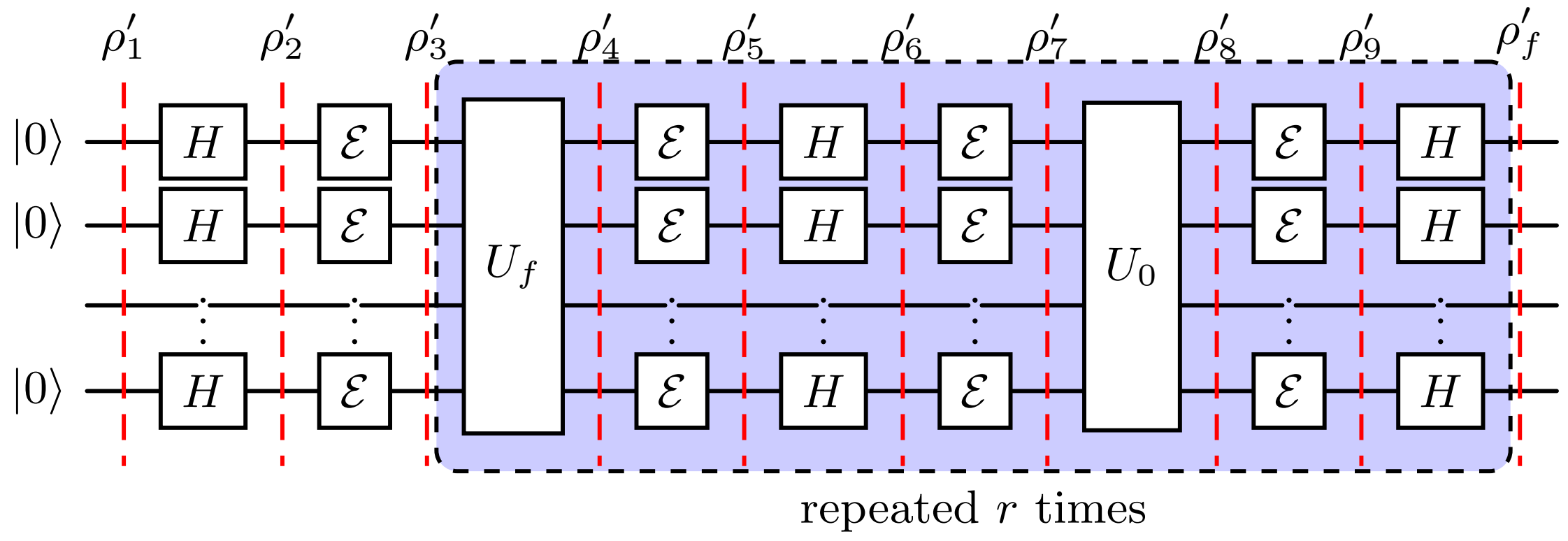

In the circuit of Fig.~\ref{fig:noisy grover} the channel of Eq.~\eqref{eq:pauli} acts on every qubit, independently, after every gate layer. One channel follows the initial Hadamard layer, and within each iteration one channel follows the oracle, one the first Hadamard layer of the diffusion step, and one the reflection $U_0$. The closing Hadamard layer of an iteration is followed directly by the oracle of the next. A circuit with $r$ iterations therefore contains $3r+1$ noise layers, and $n(3r+1)$ opportunities for an error.

Independent channels on $n$ qubits take the state out of the two-dimensional plane in which Grover's algorithm operates, and an exact treatment for all $N$ is not available. We therefore proceed in three steps. Section~\ref{sec:global} solves exactly a simpler reference model, in which the whole register is depolarized at once. Section~\ref{sec:single} works out exactly what a single error on a single qubit does to the search. Section~\ref{sec:plane} shows that for a large database the search is described by a closed set of linear equations for the part of the state that remains in the Grover plane, with rates that are simple to derive for any local channel, and Sec.~\ref{sec:solutions} solves them for bit-flip, phase-flip, depolarizing, and amplitude-damping noise.

\subsection{\label{sec:global}An exactly solvable case: global depolarizing noise}

Consider first a channel that acts on the register as a whole,
\begin{equation}
    \mathcal{E}_{\mathrm{g}}(\rho) = (1 - p_{\mathrm{g}})\, \rho + \frac{p_{\mathrm{g}}}{N}\, I ,
    \label{eq:depol}
\end{equation}
where $I$ is the $N$-dimensional identity. With probability $p_{\mathrm{g}}$ the register loses all of its information during a gate layer; with probability $1-p_{\mathrm{g}}$ nothing happens. This is the global (or total) depolarizing channel \cite{cohn2016,vrana2014}. It is not the same as $n$ independent depolarizing channels \cite{filippov2012,cohn2016}, but it is the natural idealization of them, and it has one property that makes it exactly solvable: the identity is invariant under conjugation by any unitary, $UIU^\dagger=I$, so that
\begin{equation}
    \mathcal{E}_{\mathrm{g}}(U\rho U^\dagger) = U\,\mathcal{E}_{\mathrm{g}}(\rho)\,U^\dagger .
    \label{eq:commute}
\end{equation}
The channel commutes with every gate in the circuit.

We trace one iteration. The register starts in $\rho_1' = \ket{0}^{\otimes n}\bra{0}^{\otimes n}$, and the Hadamard layer produces $\rho_2' = \ket{s}\bra{s} = \rho_2$. The first channel mixes this state with the maximally mixed state,
\begin{equation}
\rho_3' = (1 - p_{\mathrm{g}})\,\rho_2 + \frac{p_{\mathrm{g}}}{N}\,I .
\end{equation}
The populations of $\ket{w}$ and $\ket{w^\perp}$ move toward $1/N$ and the coherence between them shrinks by the factor $1-p_{\mathrm{g}}$. The oracle, Eq.~\eqref{eq:oracle}, flips the sign of the coherence and, by Eq.~\eqref{eq:commute}, passes through the mixed part unchanged,
\begin{equation}
\rho_4' = U_f\,\rho_3'\,U_f = (1-p_{\mathrm{g}})\,U_f\rho_2U_f + \frac{p_{\mathrm{g}}}{N}\,I .
\end{equation}
The second channel gives
\begin{equation}
\begin{aligned}
    \rho_5' &= (1 - p_{\mathrm{g}})\rho_4' + \frac{p_{\mathrm{g}}}{N}I \\
            &= (1-p_{\mathrm{g}})^2\,U_f\rho_2U_f + \frac{1-(1-p_{\mathrm{g}})^2}{N}\,I .
\end{aligned}
\label{eq:rho5}
\end{equation}
The diffusion step is a Hadamard layer, the reflection $U_0$, and a second Hadamard layer, with a channel after each of the first two. Every unitary leaves $I$ invariant, so the mixed parts simply add up:
\begin{align}
    \rho_6' &= H^{\otimes n}\rho_5'H^{\otimes n}, \\
    \rho_7' &= (1 - p_{\mathrm{g}})\rho_6' + \frac{p_{\mathrm{g}}}{N}I, \\
    \rho_8' &= U_0\,\rho_7'\,U_0 , \\
    \rho_9' &= (1 - p_{\mathrm{g}})\rho_8' + \frac{p_{\mathrm{g}}}{N}I , \\
    \rho_f' &= H^{\otimes n}\rho_9'H^{\otimes n} .
\end{align}
Collecting the factors, and using $V = H^{\otimes n}U_0H^{\otimes n}$ and $G=VU_f$,
\begin{equation}
    \rho_f' = (1-p_{\mathrm{g}})^4\, G\rho_2 G^{\dagger} + \frac{1-(1-p_{\mathrm{g}})^4}{N}\, I .
    \label{eq:rhof_compact}
\end{equation}
The state after one noisy iteration is the noiseless state of Eq.~\eqref{eq:rhof_noiseless}, scaled by $(1-p_{\mathrm{g}})^4$, plus a fully mixed remainder that carries the missing weight. The four channels have not changed the rotation angle. They have only reduced the weight of the rotated state.

The same pattern continues. Since every channel commutes with every gate, the $3r+1$ channels of an $r$-iteration circuit can be moved to the end of the circuit and applied together. Each one multiplies the coherent part by $1-p_{\mathrm{g}}$ and adds the missing weight to $I/N$. The state after $r$ noisy iterations is therefore
\begin{equation}
    \rho_r' = (1-p_{\mathrm{g}})^{3r+1}\, G^{r}\rho_2 G^{\dagger r} + \frac{1-(1-p_{\mathrm{g}})^{3r+1}}{N}\, I ,
    \label{eq:rhor}
\end{equation}
where $G^r\rho_2G^{\dagger r} = \ket{\psi_r}\bra{\psi_r}$ is the noiseless state of Eq.~\eqref{eq:psir}. The probability of measuring the marked state is the $\ket{w}$ population of Eq.~\eqref{eq:rhor},
\begin{equation}
\begin{aligned}
    P_r(p_{\mathrm{g}}) ={}& (1-p_{\mathrm{g}})^{3r+1}\sin^2[(2r+1)\alpha] \\
    &+ \frac{1-(1-p_{\mathrm{g}})^{3r+1}}{N} .
\end{aligned}
    \label{eq:Pr}
\end{equation}
This is the central result of this section. It reduces to Eq.~\eqref{eq:P0} at $p_{\mathrm{g}}=0$, and for any $p_{\mathrm{g}}$ it lies between the noiseless value and the random-guess value $1/N$.

Equation~\eqref{eq:Pr} is a damped oscillation in $r$. The oscillation is the Grover rotation, with the same period as in the noiseless case. The damping factor $(1-p_{\mathrm{g}})^{3r+1}$ is the probability that the register has survived $3r+1$ channels without being depolarized. Depolarizing noise therefore does not slow the rotation. It removes weight from the rotating state at a fixed rate per gate layer. Two consequences follow.

First, the maximum success probability drops below one. Its size is controlled by the product $p_{\mathrm{g}}\sqrt{N}$, because the rotation needs about $\pi\sqrt{N}/4$ iterations to complete and each iteration costs three channels. Evaluating Eq.~\eqref{eq:Pr} at $r=r_0$ for small $p_{\mathrm{g}}$ gives
\begin{equation}
    P_{r_0}(p_{\mathrm{g}}) \approx (1-p_{\mathrm{g}})^{3r_0+1} \approx e^{-3\pi p_{\mathrm{g}}\sqrt{N}/4},
    \quad p_{\mathrm{g}}\sqrt{N}\ll 1 .
    \label{eq:Pweak}
\end{equation}
A database that is four times larger needs twice as many iterations, so it tolerates only half the noise per gate layer for the same loss.

Second, the optimal number of iterations moves earlier. Setting $dP_r/dr=0$ in Eq.~\eqref{eq:Pr} with $\lambda\equiv 1-p_{\mathrm{g}}$ gives the stationarity condition
\begin{equation}
    2\alpha\,\sin[2(2r+1)\alpha] = 3\ln\!\frac{1}{\lambda}\,\Big\{\sin^2[(2r+1)\alpha]-\frac{1}{N}\Big\},
    \label{eq:stationary}
\end{equation}
whose left side is the gain from one more rotation step and whose right side is the loss from three more channels. The optimal count $r^{*}$ is the integer near the root of Eq.~\eqref{eq:stationary} in the first Grover period, $1\le r\le \pi/(4\alpha)$; beyond it the damping only continues. For weak noise the root lies close to $r_0$. Writing $(2r+1)\alpha = \pi/2-\delta$ with $\delta$ small, and using $\ln(1/\lambda)\approx p_{\mathrm{g}}$ and $\alpha^2\approx 1/N$,
\begin{equation}
    r^{*} \approx r_0 - \frac{3p_{\mathrm{g}}\,(N-1)}{8},
    \qquad p_{\mathrm{g}}\sqrt{N}\ll 1 .
    \label{eq:rweak}
\end{equation}
The shift is proportional to $p_{\mathrm{g}}N$, so at fixed $p_{\mathrm{g}}$ it grows faster than $r_0$ itself. Once $p_{\mathrm{g}}\sqrt{N}$ is of order one, the optimum is no longer a small correction to $r_0$. Once $p_{\mathrm{g}}\sqrt{N}\gg1$ (with $p_{\mathrm{g}}$ still small) the coherent part is damped away long before the rotation completes. In that regime $\sin[(2r+1)\alpha]\approx(2r+1)\alpha$, and maximizing $\lambda^{3r+1}(2r+1)^2$ gives
\begin{equation}
    r^{*} \approx \frac{2}{3p_{\mathrm{g}}} - \frac{1}{2},
    \quad
    P_{r^*} \approx \frac{16}{9e^{2}}\,\frac{1}{p_{\mathrm{g}}^{2}N} ,
    \quad p_{\mathrm{g}}\sqrt{N}\gg 1 .
    \label{eq:rstrong}
\end{equation}
The optimal iteration count then no longer depends on $N$ at all. It is set by the number of gate layers the register can survive, and the success probability is only a constant factor above the random-guess value $1/N$.

A run with $r$ iterations makes $r$ oracle queries and succeeds with probability $P_r$. If the run is repeated until the measured item passes a check, the expected number of oracle queries is
\begin{equation}
    Q(r) = \frac{r}{P_r} ,
    \qquad
    Q^{*} = \min_{1\le r\le \pi/(4\alpha)} Q(r) .
    \label{eq:Q}
\end{equation}
(The check itself adds one query per run and is omitted; it does not change any scaling.) $Q^{*}$ is the natural measure of the cost of the noisy search, and $Q_0$ is its noiseless value. $Q_0$ is slightly below $r_0$, because it is cheaper to stop a little before the peak of $P^{(0)}_r$ and repeat occasionally than to run to the peak every time \cite{boyer1998}. In the weak-noise regime Eq.~\eqref{eq:Pweak} gives $Q^{*}/Q_0\approx \exp(3\pi p_{\mathrm{g}}\sqrt{N}/4)$ to leading order, so the overhead depends on $p_{\mathrm{g}}$ and $N$ only through the product $p_{\mathrm{g}}\sqrt{N}$. In the strong-noise regime Eq.~\eqref{eq:rstrong} gives a cost proportional to $p_{\mathrm{g}}N$; minimizing Eq.~\eqref{eq:Q} directly rather than using $r^*$ yields the prefactor
\begin{equation}
    Q^{*} \approx \frac{3e}{4}\,p_{\mathrm{g}}N \approx 2\,p_{\mathrm{g}}N ,
    \qquad p_{\mathrm{g}}\sqrt{N}\gg 1 .
    \label{eq:Qstrong}
\end{equation}
The cost is then linear in $N$, as for a classical search, and the quantum algorithm retains only a constant factor of about $4p_{\mathrm{g}}$ over a classical scan of half the database. This is the scaling allowed by the lower bounds of Refs.~\cite{vrana2014,rosmanis2023}, which show that any search facing an oracle that depolarizes with a fixed probability needs a number of queries of order $pN$. Equation~\eqref{eq:Qstrong} shows that Grover's algorithm, stopped at the right point and repeated, attains it with a small prefactor.

\begin{figure}[t]
    \centering
    \includegraphics[width=\linewidth]{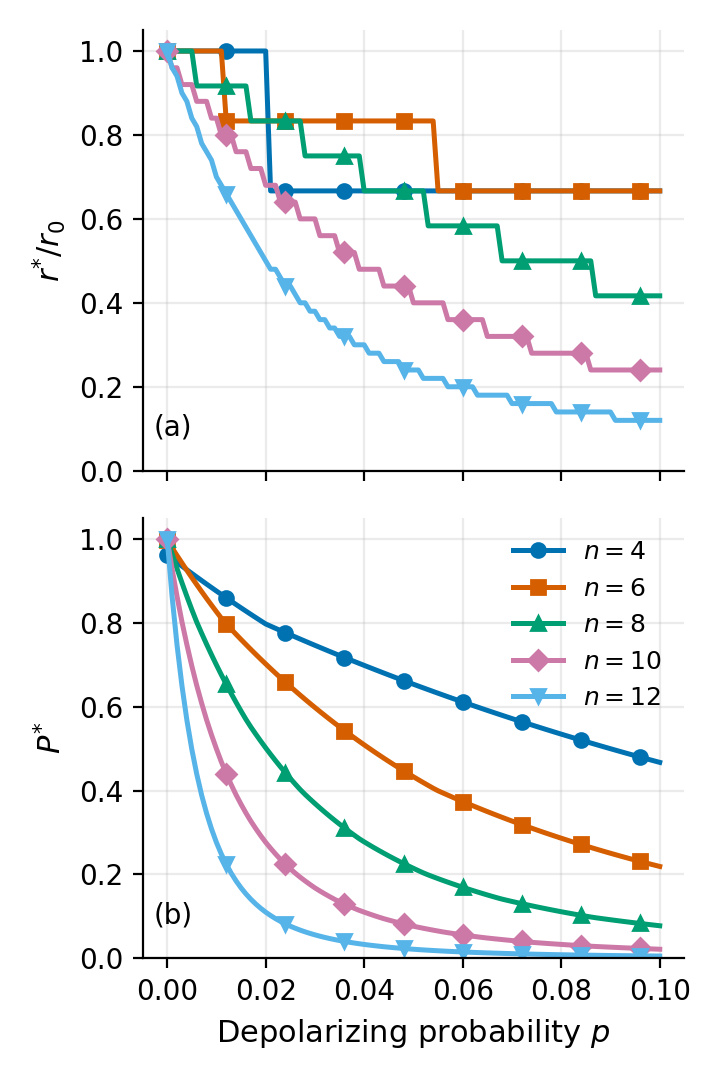}
    \caption{Global depolarizing noise. (a) Optimal iteration count $r^{*}$ relative to the noiseless value $r_0$, and (b) the maximum success probability $P^{*}$, as functions of $p_{\mathrm{g}}$ for several register sizes $n$, from Eq.~\eqref{eq:Pr}. The optimal count never exceeds $r_0$ and drops the faster the larger the register; the steps in (a) reflect the integer nature of $r^{*}$. Plotting code written with AI assistance (see Acknowledgments).}
    \label{fig:r vs p}
\end{figure}

\begin{figure}[t]
    \centering
    \includegraphics[width=\linewidth]{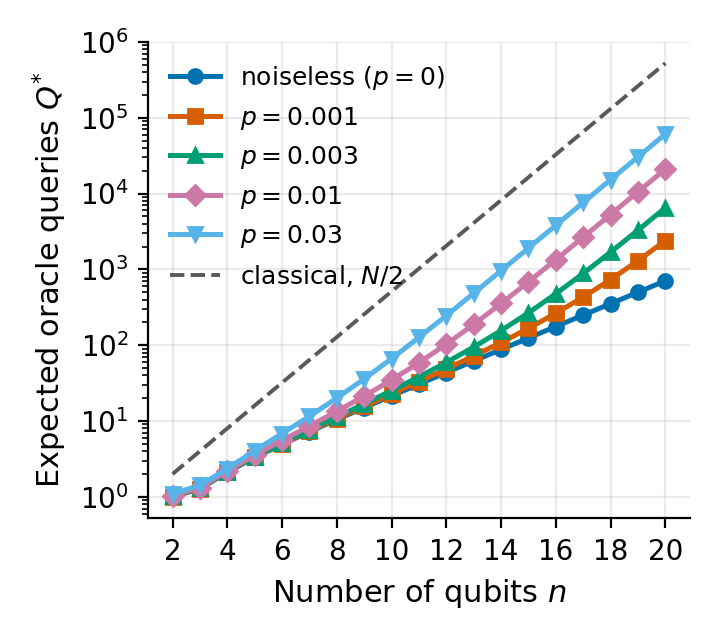}
    \caption{Global depolarizing noise. Expected number of oracle queries $Q^{*}$, Eq.~\eqref{eq:Q}, as a function of the register size $n$ for several values of $p_{\mathrm{g}}$. The dashed line is the classical cost $N/2$. Each noisy curve follows the noiseless $\sqrt{N}$ line while $p_{\mathrm{g}}\sqrt{N}$ is small and then bends over to the classical slope, $Q^{*}\approx 2p_{\mathrm{g}}N$ from Eq.~\eqref{eq:Qstrong}. Plotting code written with AI assistance (see Acknowledgments).}
    \label{fig:r vs n}
\end{figure}

\begin{figure}[t]
    \centering
    \includegraphics[width=\linewidth]{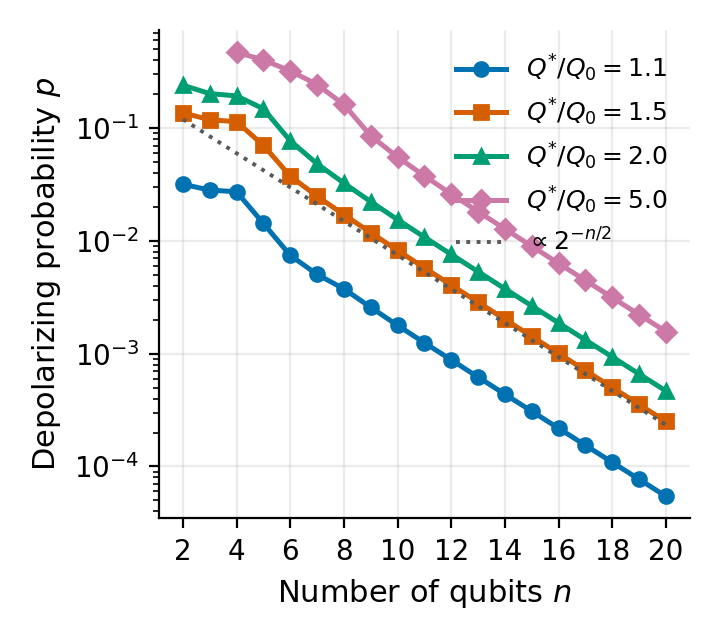}
    \caption{Global depolarizing noise. Value of $p_{\mathrm{g}}$ that produces a fixed cost overhead $Q^{*}/Q_0$, as a function of the register size $n$. The dotted line is proportional to $2^{-n/2}=1/\sqrt{N}$. For each overhead the tolerable noise decreases as $1/\sqrt{N}$ once the register exceeds a few qubits. Plotting code written with AI assistance (see Acknowledgments).}
    \label{fig:p vs n}
\end{figure}

Figures~\ref{fig:r vs p} to \ref{fig:p vs n} show these results. Figure~\ref{fig:r vs p} shows that noise moves the optimal iteration count down, not up, and the faster the larger the register, in agreement with Eq.~\eqref{eq:rweak}. The success probability falls correspondingly. Figure~\ref{fig:r vs n} shows the cost of the search as the register grows. On the logarithmic scale the noiseless cost is a straight line whose slope is the signature of the $\sqrt{N}$ scaling. Each noisy curve tracks this line at first and then bends over to a line of twice the slope, parallel to the classical cost. The bend occurs where $p_{\mathrm{g}}\sqrt{N}$ reaches about one half. Beyond the bend the search still beats a classical scan, but only by the constant factor of Eq.~\eqref{eq:Qstrong}. For a fixed noise level, therefore, Grover's algorithm has a maximum useful database size, set by $p_{\mathrm{g}}\sqrt{N}\sim 1$, that is, by $n_{\max}\approx 2\log_2(1/p_{\mathrm{g}})$. Figure~\ref{fig:p vs n} inverts the question and asks what noise level can be tolerated if the search is to cost at most a given factor more than in the noiseless case. For every overhead the answer falls as $2^{-n/2}$ once $n$ exceeds a few qubits: adding two qubits quadruples $N$ and halves the tolerable $p_{\mathrm{g}}$.

\subsection{\label{sec:single}Local noise: what a single error does}

We now return to independent channels on the individual qubits. (A related but different situation, in which the same Pauli error is applied jointly to a fixed set of qubits, was studied in Ref.~\cite{mandal2023}; there, for bit-flip and phase-flip errors, the success probability turns out not to depend on the size of the set. Here the errors on different qubits are independent, which is why the number of qubits will enter every rate.) The key to understanding them is the geometry of Sec.~\ref{sec:noiseless}. The Hilbert space of the register splits into the plane spanned by $\ket{w^\perp}$ and $\ket{w}$ and its orthogonal complement. The Grover operator rotates the plane by $2\alpha$. On the complement it does something much simpler: the oracle leaves every state orthogonal to $\ket{w}$ unchanged, and the diffusion operator reverses the sign of every state orthogonal to $\ket{s}$, so
\begin{equation}
    G\ket{\chi} = -\ket{\chi}
    \qquad \text{for every } \ket{\chi} \perp \ket{w^\perp},\ket{w} .
    \label{eq:frozen}
\end{equation}
Amplitude that leaves the plane is therefore frozen. It is never amplified, and the ideal gates never bring it back. Only a later error can, and we will see that this is rare.

A Pauli error on one qubit sends part of the state back into the plane and part of it out. The in-plane part is all that matters for the final success probability, because the frozen part has no overlap with $\ket{w}$. Taking $w=0\cdots0$ without loss of generality, and writing $\Pi=\ket{w^\perp}\bra{w^\perp}+\ket{w}\bra{w}$ for the projector onto the plane, the in-plane parts of the three Pauli operators on qubit $k$ are
\begin{equation}
\begin{aligned}
    \Pi Z_k \Pi &= \ket{w}\bra{w} - \frac{1}{N-1}\,\ket{w^\perp}\bra{w^\perp}, \\
    \Pi X_k \Pi &= \frac{N-2}{N-1}\,\ket{w^\perp}\bra{w^\perp}
      + \frac{\ket{w}\bra{w^\perp} + \ket{w^\perp}\bra{w}}{\sqrt{N-1}}, \\
    \Pi Y_k \Pi &= \frac{i}{\sqrt{N-1}}\big(\ket{w^\perp}\bra{w} - \ket{w}\bra{w^\perp}\big),
\end{aligned}
\label{eq:pauli_plane}
\end{equation}
the same for every qubit $k$. For a large database the terms of order $1/\sqrt{N}$ are negligible and the picture is simple. A phase flip keeps the marked component of the state and throws the unmarked component out of the plane: $Z_k\ket{w}=\pm\ket{w}$, while $Z_k\ket{w^\perp}$ is a superposition of unmarked states with scrambled signs, almost orthogonal to $\ket{w^\perp}$. A bit flip does the opposite. It keeps the unmarked component, since flipping one bit of an unmarked state almost always gives another unmarked state, and throws the marked component out, since $X_k\ket{w}$ is an unmarked state. A $Y$ error, which is a bit flip and a phase flip together, throws both components out. In every case the coherence between $\ket{w^\perp}$ and $\ket{w}$ is lost, because one of the two components has left the plane.

What this means for the search follows from the rotation picture. Suppose a single error occurs when the in-plane state is at angle $\theta$ from $\ket{w^\perp}$, and that the remaining iterations would have carried an unperturbed state the rest of the way to $\pi/2$. After a phase flip the in-plane state is $\sin\theta\ket{w}$, which sits at angle $\pi/2$ and is carried by the remaining rotation to angle $\pi-\theta$, so the final success probability is
\begin{equation}
    P^{(Z)} \approx \sin^2\theta\;\sin^2(\pi-\theta) = \sin^4\theta .
    \label{eq:PZ}
\end{equation}
After a bit flip the in-plane state is $\cos\theta\ket{w^\perp}$, at angle zero, carried to angle $\pi/2-\theta$, so
\begin{equation}
    P^{(X)} \approx \cos^2\theta\;\cos^2\theta = \cos^4\theta .
    \label{eq:PX}
\end{equation}
After a $Y$ error essentially nothing is left in the plane and $P^{(Y)}\approx 1/N$. Equations~\eqref{eq:PZ} and \eqref{eq:PX} say that the two basic errors are damaging at opposite times. A phase flip is worst early in the search, when almost all the amplitude still sits on unmarked states; a single phase flip on any one qubit right after the initial Hadamard layer leaves a success probability of order $1/N$ and ruins the search outright. A bit flip is worst late, when the amplitude has gathered on the marked state; a bit flip at the very beginning costs nothing at all, because $X_k\ket{s}=\ket{s}$. A $Y$ error is fatal at any time. Averaged over the course of a full search, with $\theta$ sweeping uniformly from $0$ to $\pi/2$, a phase flip or a bit flip costs on average a fraction
\begin{equation}
    c_X = c_Z = \Big\langle 1-\sin^4\theta\Big\rangle = \frac{5}{8}
    \label{eq:c58}
\end{equation}
of the success probability, and a $Y$ error costs $c_Y=1$. The value $5/8$ for phase flips agrees with Azuma's perturbative analysis of $\sigma_z$ errors on the individual qubits \cite{azuma2002}, which finds the same loss of success probability per expected error.

One more point concerns where in the circuit the error occurs. The diffusion operator is $H^{\otimes n}U_0H^{\otimes n}$, and $HXH=Z$, $HZH=X$, $HYH=-Y$. An error that occurs between the two Hadamard layers is therefore equivalent to the conjugate error applied outside them: a bit flip inside the diffusion step acts on the register as a phase flip, and vice versa, while a $Y$ error stays a $Y$ error. Of the three noise layers in each iteration of Fig.~\ref{fig:noisy grover}, one sees the error as it is and two see it conjugated. Bit-flip noise in the physical circuit thus produces one bit-flip-like and two phase-flip-like error events per iteration, and phase-flip noise the reverse.

\subsection{\label{sec:plane}The Grover plane as a noisy qubit}

The single-error picture extends to many errors, and the extension is exact in the limit of a large database. The reason is Eq.~\eqref{eq:frozen}. Once amplitude has left the plane it stays out, so it can be discarded, and the search is described by the part of the density matrix that lives in the plane,
\begin{equation}
    \rho_\Pi = \Pi\rho\,\Pi =
    \begin{pmatrix} A & C \\ C & B \end{pmatrix}
    \quad\text{in the basis } \{\ket{w^\perp},\ket{w}\},
    \label{eq:rhoPi}
\end{equation}
with $B$ the success probability, $A$ the weight on unmarked states that is still available for amplification, and $C$ the coherence that the Grover rotation converts from one into the other. Frozen amplitude can return to the plane only if a later error happens to undo an earlier one on the same qubit, which is a fraction of order $1/n$ of all two-error events, and we neglect it. We also drop the terms of order $1/\sqrt{N}$ in Eq.~\eqref{eq:pauli_plane}. Within these approximations $\rho_\Pi$ obeys a closed equation of motion: the ideal gates act on it as the reflections of Sec.~\ref{sec:noiseless}, and each noise layer acts on it as follows.

Consider a layer of the Pauli channel of Eq.~\eqref{eq:pauli} on all $n$ qubits, seen as it is. The unmarked weight $A$ survives the layer if no qubit suffers a phase flip or a $Y$ error, which happens with probability $(1-p_z-p_y)^n$; the marked weight $B$ survives if no qubit suffers a bit flip or a $Y$ error, with probability $(1-p_x-p_y)^n$; and the coherence $C$ survives only if no qubit suffers any error at all, with probability $(1-p)^n$. A layer seen through the Hadamard conjugation has $p_x$ and $p_z$ exchanged. For small $p$ the three survival probabilities are $1-n(p_z+p_y)$, $1-n(p_x+p_y)$, and $1-np$. Adding up the three layers of an iteration, one as it is and two conjugated, gives the decay rates per iteration
\begin{equation}
\begin{aligned}
    \gamma_A &= n\,(p_z + 2p_x + 3p_y), \\
    \gamma_B &= n\,(p_x + 2p_z + 3p_y), \\
    \gamma_C &= 3n\,(p_x+p_y+p_z) = 3np .
\end{aligned}
\label{eq:rates_pauli}
\end{equation}
The rotation, in the same units, advances the angle by $2\alpha$ per iteration. For a large database $\alpha$ is small and the iterations can be treated as a continuous time $t$ counted in iterations. The reflections of Sec.~\ref{sec:noiseless} then combine into a rotation at the angular rate $\omega=2\alpha$, under which $\dot A=-2\omega C$, $\dot B=2\omega C$, $\dot C=\omega(A-B)$, and the noise adds the decay terms. The result is the set of linear equations
\begin{equation}
\begin{aligned}
    \dot A &= -2\omega\, C - \gamma_A A, \\
    \dot B &= \phantom{-}2\omega\, C - \gamma_B B, \\
    \dot C &= \omega\,(A-B) - \gamma_C\, C ,
\end{aligned}
\qquad \omega = 2\alpha ,
\label{eq:plane_ode}
\end{equation}
with the initial condition $A=\cos^2\alpha$, $B=\sin^2\alpha$, $C=\sin\alpha\cos\alpha$ of Eq.~\eqref{eq:alpha}. We call these the Grover-plane equations. They are the Bloch equations of a two-level system driven at a constant rate, with different decay rates for the two populations and for the coherence. They say that the Grover plane behaves as a qubit that is rotated by the algorithm and, at the same time, leaks weight out of the plane at the rates $\gamma_A$ and $\gamma_B$ and loses its coherence at the rate $\gamma_C$. The global channel of Sec.~\ref{sec:global} is the special case $\gamma_A=\gamma_B=\gamma_C=3p_{\mathrm{g}}$, for which Eq.~\eqref{eq:plane_ode} is solved by Eq.~\eqref{eq:Pr} in the continuum limit. Local noise differs from it in one respect only: it destroys the coherence faster than it drains the weight, because an error on a single qubit removes one component from the plane but not the other. Table~\ref{tab:rates} lists the rates for the channels of interest.

\begin{table}[t]
\caption{\label{tab:rates}Decay rates per Grover iteration in the Grover-plane equations \eqref{eq:plane_ode}, for noise of strength $p$ (Pauli channels) or $\gamma$ (amplitude damping) per qubit and gate layer, in the circuit of Fig.~\ref{fig:noisy grover}. For amplitude damping, $h$ is the number of ones in the binary label of the marked item.}
\begin{ruledtabular}
\setlength{\tabcolsep}{3pt}
\begin{tabular}{lccc}
Channel & $\gamma_A$ & $\gamma_B$ & $\gamma_C$ \\
\hline
global depolarizing & $3p_{\mathrm{g}}$ & $3p_{\mathrm{g}}$ & $3p_{\mathrm{g}}$ \\
bit flip & $2np$ & $np$ & $3np$ \\
phase flip & $np$ & $2np$ & $3np$ \\
depolarizing & $2np$ & $2np$ & $3np$ \\
amplitude damping & $\tfrac{n}{4}\gamma$ & $(\tfrac{n}{2}+h)\gamma$ & $(\tfrac{3n}{4}+\tfrac{h}{2})\gamma$ \\
\end{tabular}
\end{ruledtabular}
\end{table}

Amplitude damping is not a Pauli channel, but it is handled in the same way. On one qubit it has the Kraus operators $K_0=\mathrm{diag}(1,\sqrt{1-\gamma})$ and $K_1=\sqrt{\gamma}\,\ket{0}\bra{1}$: with probability $\gamma$ an excited qubit decays to $\ket{0}$, and otherwise its excited amplitude is reduced. Since the channel singles out $\ket{0}$, the result depends on the marked item, through the number $h$ of ones in its binary label. To first order in $\gamma$, $K_0^{\otimes n}$ reduces the marked amplitude by $h\gamma/2$ and the unmarked amplitude by $n\gamma/4$, because an average unmarked state has half of its qubits excited; the decay events $K_1$ on a marked qubit send $\ket{w}$ out of the plane, while on $\ket{w^\perp}$ they leave a fraction $\gamma/4$ per qubit inside it but without coherence. Between the two Hadamard layers the channel damps toward $\ket{+}$ on every qubit, which is to say toward $\ket{s}$: there it leaves $\ket{w^\perp}$ alone and reduces the marked weight by $\gamma/4$ per qubit. Collecting the three layers gives the last row of Table~\ref{tab:rates}. The marked weight always decays faster than the unmarked weight under amplitude damping, and the worst case is a marked item whose label is all ones.

\subsection{\label{sec:solutions}Solutions and the noise budget}

Equation~\eqref{eq:plane_ode} is a linear system with constant coefficients, so its solution is a sum of three exponentials whose exponents are the roots of a cubic. The structure is clearest in the variables $S=A+B$, the total weight left in the plane, and $D=A-B$, the population difference. With $\bar\gamma=(\gamma_A+\gamma_B)/2$, $\delta=(\gamma_A-\gamma_B)/2$, and $\kappa=(\gamma_C-\bar\gamma)/2$,
\begin{equation}
\begin{aligned}
    \dot S &= -\bar\gamma\, S - \delta\, D, \\
    \dot D &= -\bar\gamma\, D - \delta\, S - 4\omega C, \\
    \dot C &= \omega D - (\bar\gamma + 2\kappa)\, C .
\end{aligned}
\label{eq:SD}
\end{equation}
The weight $S$ decays at the mean rate $\bar\gamma$. The pair $(D,C)$ is a damped oscillator: it carries the Grover oscillation at the rate $2\omega$ and decays at the rate $\bar\gamma+\kappa$, which exceeds the decay of the weight by the extra dephasing $\kappa$. The asymmetry $\delta$ couples the two.

When the two populations decay at the same rate, $\delta=0$, the system separates and can be solved in closed form. Writing $\gamma$ for the common rate, the weight is $S=e^{-\gamma t}$, and the oscillator gives, for $\kappa\ll 2\omega$,
\begin{equation}
    P_r \simeq \frac{e^{-\gamma r}}{2}\Big\{1 - e^{-\kappa r}\Big[\cos 2(2r+1)\alpha + \frac{\kappa}{4\alpha}\sin 4\alpha r\Big]\Big\} .
    \label{eq:Plocal}
\end{equation}
The first term is half the weight remaining in the plane. The second is the Grover oscillation, which decays faster than the weight by the factor $e^{-\kappa r}$ and acquires a small phase lag through the term in $\kappa/4\alpha$. At $\kappa=0$ Eq.~\eqref{eq:Plocal} reduces to the continuum form of the global result, Eq.~\eqref{eq:Pr}. Equation~\eqref{eq:Plocal} applies directly to the depolarizing channel, with $\gamma=2np$ and $\kappa=np/2$ from Table~\ref{tab:rates}, and to any Pauli channel with $p_x=p_z$.

For the bit-flip and phase-flip channels, and for amplitude damping, the two populations decay at different rates and $\delta\neq0$. Equation~\eqref{eq:SD} is still exactly solvable, through the cubic
\begin{equation}
    \mu^3 + 2\kappa\,\mu^2 + (4\omega^2-\delta^2)\,\mu - 2\kappa\,\delta^2 = 0
    \label{eq:cubic}
\end{equation}
for the exponents $-\bar\gamma+\mu$, but the roots are not illuminating. What matters is the size of the correction. The asymmetry enters only through its interplay with the rotation, in the ratio $\delta/\omega$, and the corrections it produces to Eq.~\eqref{eq:Plocal} with $\gamma\to\bar\gamma$ are of first order in that ratio and remain small at the noiseless optimum, where the two populations have exchanged their roles. For the bit-flip and phase-flip channels, which have $\bar\gamma=3np/2$ and $\kappa=3np/4$, the corrections at the optimal iteration count amount to a few percent of the success probability, with opposite signs for the two channels: bit-flip noise is slightly less damaging than phase-flip noise, and both are less damaging than depolarizing noise at the same total error probability. For amplitude damping the asymmetry is larger, $\delta=-(n/4+h)\gamma/2$, and the corrections can reach several percent.

Everything of practical importance follows from the leading behavior, which is common to all local channels. At the noiseless optimum, where the cosine in Eq.~\eqref{eq:Plocal} is $-1$,
\begin{equation}
    P_{r_0} \simeq \frac{e^{-\bar\gamma r_0}}{2}\big(1 + e^{-\kappa r_0}\big)
    \approx \exp\!\Big[-\big(\bar\gamma+\tfrac{\kappa}{2}\big)\frac{\pi\sqrt{N}}{4}\Big] ,
    \label{eq:Plocal_r0}
\end{equation}
the last form holding to first order in the rates. The combination $\bar\gamma+\kappa/2$ is the effective decay rate of the search. From Table~\ref{tab:rates} it is $3\cdot\tfrac{5}{8}\,np$ for the bit-flip and phase-flip channels, $3\cdot\tfrac{3}{4}\,np$ for the depolarizing channel, and $\tfrac{15}{32}n\gamma+\tfrac{1}{2}h\gamma$ for amplitude damping. The first two reproduce the average costs per error of Sec.~\ref{sec:single}: to first order, local noise of strength $p$ per qubit acts like a global channel of rate $c\,n\,p$ with $c=5/8$ for a bit or phase flip and $c=3/4$ for depolarizing noise. Beyond first order the local channels differ from the global one in the way the plane decoheres, Eq.~\eqref{eq:Plocal}, but not in the scaling. All the conclusions of Sec.~\ref{sec:global} therefore carry over with $p_{\mathrm{g}}$ replaced by $cnp$. The maximum useful register size is set by $cnp\sqrt{N}\sim1$, and the noise budget per qubit and gate layer is
\begin{equation}
    p \lesssim \frac{1}{c\,n\sqrt{N}} .
    \label{eq:budget}
\end{equation}
The extra factor of $n$, compared with the global channel, simply counts the qubits that can each go wrong. It makes the budget tighter by one power of the register size on top of the $\sqrt{N}$ from the circuit depth. For local depolarizing noise, Eq.~\eqref{eq:budget} has the same form as the threshold of order $1/(\sqrt{N}\log_2 N)$ found by Cohn and co-workers \cite{cohn2016}, since $\log_2 N = n$. The present derivation identifies the constant, explains it through the single-error picture of Sec.~\ref{sec:single}, and extends it to the other channels of Table~\ref{tab:rates}.

\begin{figure}[t]
    \centering
    \includegraphics[width=\linewidth]{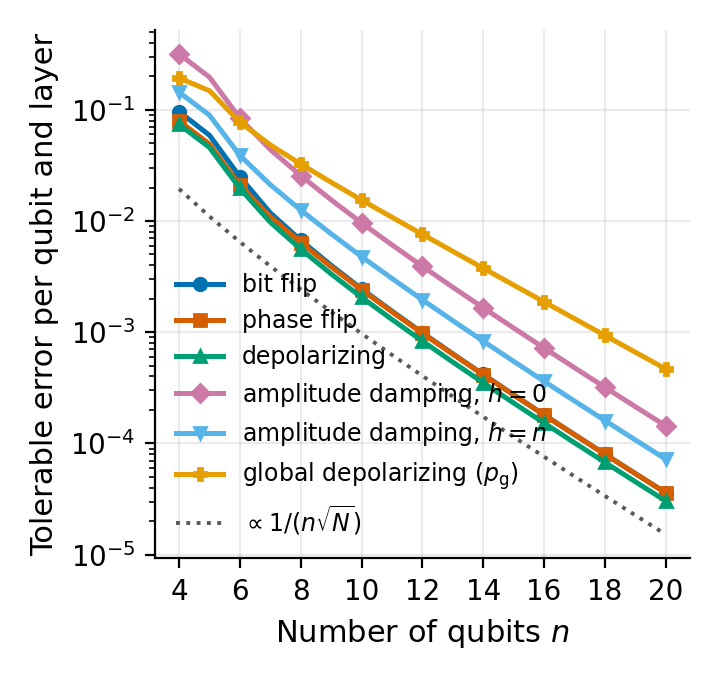}
    \caption{Error per qubit and gate layer that doubles the expected query cost, $Q^{*}/Q_0=2$, as a function of the register size $n$, from the Grover-plane equations \eqref{eq:plane_ode} with the rates of Table~\ref{tab:rates}. For amplitude damping the best case ($h=0$, marked label all zeros) and the worst case ($h=n$, all ones) are shown. The global depolarizing channel of Fig.~\ref{fig:p vs n} is included for comparison; its rate $p_{\mathrm{g}}$ refers to the whole register. The dotted line is proportional to $1/(n\sqrt{N})$, Eq.~\eqref{eq:budget}. Plotting code written with AI assistance (see Acknowledgments).}
    \label{fig:budget plane}
\end{figure}

Figure~\ref{fig:budget plane} shows the budget for all the channels of Table~\ref{tab:rates}, obtained from Eq.~\eqref{eq:plane_ode} and the cost of Eq.~\eqref{eq:Q}. Beyond about six qubits every local curve falls parallel to the $1/(n\sqrt{N})$ line, one power of $n$ steeper than the global channel. The ordering follows the effective rates: depolarizing noise is the least tolerable of the Pauli channels, bit-flip and phase-flip noise are nearly equivalent, and amplitude damping is tolerated at a several times higher rate, because a decay event on an unmarked state leaves part of the weight in the plane and because the channel only acts on excited qubits. Even in the worst case of a marked label of all ones, amplitude damping is less harmful than a bit flip of the same probability, since it damps rather than flips. In every case the lesson is the same: each additional qubit costs a factor of about $\sqrt2$ in the tolerable error from the longer circuit and a further factor $(n+1)/n$ from the extra qubit that can fail.

\section{\label{sec:rot}Grover's algorithm under noisy rotations}

\subsection{Noise model}

Depolarizing noise randomizes the state. A second, quite different kind of error leaves the state pure but moves it to the wrong place. If the oracle or the diffusion operator is implemented with a slightly wrong phase or a slightly wrong rotation axis, each Grover iteration rotates the state by an angle that differs a little from $2\alpha$. Unlike depolarizing noise, such coherent errors can have the same sign in every iteration, for instance when they come from a miscalibrated gate, and then they add up linearly with the circuit depth \cite{preskill2018,wallman2016}.

We model this as a bounded angular error. In the plane of $\ket{w^\perp}$ and $\ket{w}$, an ideal oracle reflects the state about the $\ket{w^\perp}$ axis and an ideal diffusion operator reflects it about $\ket{s}$. We allow the angle of the state after the oracle to be off by at most $\delta$, and the angle after the diffusion operator to be off by a further amount of at most $\Delta$. The signs of the errors are not specified, so the state after each operation lies inside an angular interval rather than at a single angle. This is a worst-case model: it asks how far the state can wander, not how far it typically wanders.

\subsection{Error propagation}

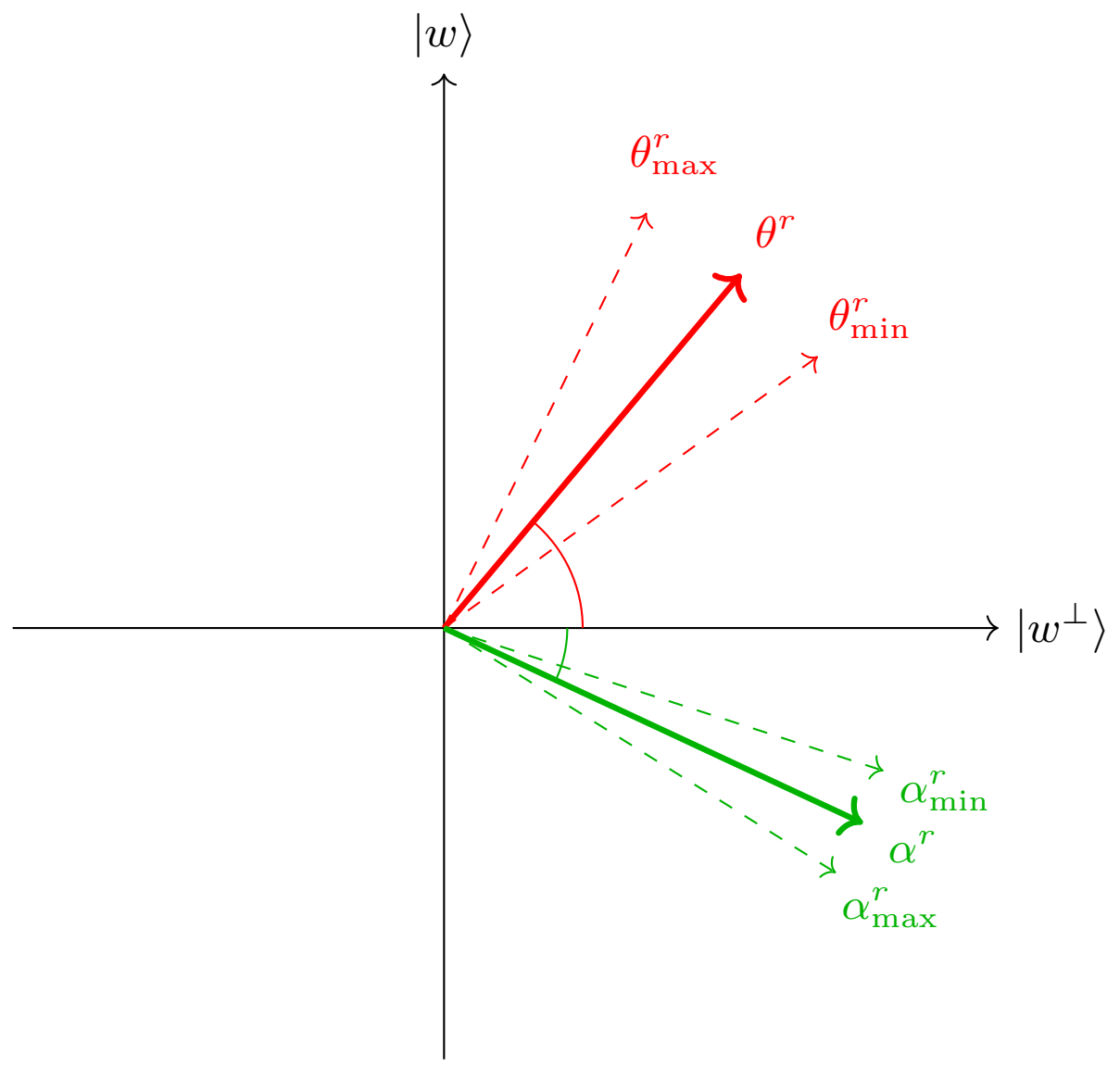
\begin{figure}[t]
\centering
\begin{tikzpicture}[scale=2.2]
\draw[->] (-1.4,0) -- (1.8,0) node[right] {$\ket{w^\perp}$};
\draw[->] (0,-1.4) -- (0,1.8) node[above] {$\ket{w}$};
\def\alphaAngle{-25}
\def\thetaAngle{-2*\alphaAngle}
\def\len{1.5}
\def\alphaMin{-18}
\def\alphaMax{-32}
\def\thetaMin{-2*\alphaMin}
\def\thetaMax{-2*\alphaMax}
\draw[very thick, red, ->]
(0,0) -- ({\len*cos(\thetaAngle)}, {\len*sin(\thetaAngle)});
\node[red] at
({1.12*\len*cos(\thetaAngle)}, {1.12*\len*sin(\thetaAngle)})
{\small $\theta^r$};
\draw[dashed, red, ->]
(0,0) -- ({\len*cos(\thetaMin)}, {\len*sin(\thetaMin)});
\node[red] at
({1.14*\len*cos(\thetaMin)}, {1.14*\len*sin(\thetaMin)})
{\small $\theta^r_{\min}$};
\draw[dashed, red, ->]
(0,0) -- ({\len*cos(\thetaMax)}, {\len*sin(\thetaMax)});
\node[red] at
({1.14*\len*cos(\thetaMax)}, {1.14*\len*sin(\thetaMax)})
{\small $\theta^r_{\max}$};
\draw[red] (0.45,0) arc (0:\thetaAngle:0.45);
\draw[very thick, green!70!black, ->]
(0,0) -- ({\len*cos(\alphaAngle)}, {\len*sin(\alphaAngle)});
\node[green!70!black] at
({1.12*\len*cos(\alphaAngle)}, {1.12*\len*sin(\alphaAngle)})
{\small $\alpha^r$};
\draw[dashed, green!70!black, ->]
(0,0) -- ({\len*cos(\alphaMin)}, {\len*sin(\alphaMin)});
\node[green!70!black] at
({1.14*\len*cos(\alphaMin)}, {1.14*\len*sin(\alphaMin)})
{\small $\alpha^r_{\min}$};
\draw[dashed, green!70!black, ->]
(0,0) -- ({\len*cos(\alphaMax)}, {\len*sin(\alphaMax)});
\node[green!70!black] at
({1.14*\len*cos(\alphaMax)}, {1.14*\len*sin(\alphaMax)})
{\small $\alpha^r_{\max}$};
\draw[green!70!black] (0.4,0) arc (0:\alphaAngle:0.4);
\end{tikzpicture}
\caption{Geometry of the noisy rotations in the plane spanned by $\ket{w^\perp}$ and $\ket{w}$. The oracle reflects the state below the $\ket{w^\perp}$ axis, to an angle of magnitude $\alpha^{r}$ (green). The diffusion operator reflects it back above the axis, to the angle $\theta^{r}$ (red). With bounded errors each angle lies inside an interval, shown by the dashed rays, and the intervals widen from one iteration to the next.}
\label{fig:noisy_angles}
\end{figure}

Figure~\ref{fig:noisy_angles} shows the geometry. Let $\theta^{r}$ denote the angle of the state, measured from $\ket{w^\perp}$, after $r$ complete iterations, with $\theta^{0}=\alpha$. The oracle of iteration $r$ reflects the state to the angle $-\theta^{r-1}$ below the axis, up to the error $\delta$. We write $\alpha^{r}$ for the magnitude of this angle (the green rays in the figure). The diffusion operator then reflects the state about $\ket{s}$, which takes an angle $-\phi$ to $2\alpha+\phi$, up to the error $\Delta$ (the red rays).

For the first iteration the bounds are
\begin{align}
&\alpha^{1}_{\min} = \alpha - \delta, &
&\alpha^{1}_{\max} = \alpha + \delta,\\
&\theta^{1}_{\min} = 3\alpha - \delta - \Delta, &
&\theta^{1}_{\max} = 3\alpha + \delta + \Delta,
\end{align}
and for the second
\begin{align}
&\alpha^{2}_{\min} = 3\alpha - 2\delta - \Delta, &
&\alpha^{2}_{\max} = 3\alpha + 2\delta + \Delta,\\
&\theta^{2}_{\min} = 5\alpha - 2\delta - 2\Delta, &
&\theta^{2}_{\max} = 5\alpha + 2\delta + 2\Delta.
\end{align}
Each oracle widens the interval by $\delta$ on either side and each diffusion operator by $\Delta$, while the center advances by $2\alpha$ per iteration. The general recursion is
\begin{align}
\alpha^{r}_{\min} &= \theta^{r-1}_{\min} - \delta = (2r-1)\alpha - r \delta - (r-1) \Delta,\\
\alpha^{r}_{\max} &= \theta^{r-1}_{\max} + \delta = (2r-1)\alpha + r \delta + (r-1) \Delta,\\
\theta^{r}_{\min} &= \alpha^{r}_{\min} + 2\alpha - \Delta = (2r+1)\alpha - r (\delta + \Delta),\\
\theta^{r}_{\max} &= \alpha^{r}_{\max} + 2\alpha + \Delta = (2r+1)\alpha + r (\delta + \Delta).
\label{eq:theta_bounds}
\end{align}
The width of the angular interval after $r$ iterations is
\begin{equation}
\theta^{r}_{\max} - \theta^{r}_{\min} = 2r (\delta + \Delta).
\label{eq:width}
\end{equation}
It grows linearly with the number of iterations. Since the algorithm needs about $\pi\sqrt{N}/4$ iterations, the accumulated uncertainty at the end of the search is of order $\sqrt{N}(\delta+\Delta)$. The same combination of noise strength and database size that governed the depolarizing case governs this one.

\subsection{Bounds on the error probability}

After $r$ iterations the state is $\cos\phi\,\ket{w^\perp}+\sin\phi\,\ket{w}$ for some angle $\phi$ in the interval $[\theta^{r}_{\min},\theta^{r}_{\max}]$. The probability of measuring an unmarked item, which we call the error probability, is
\begin{equation}
P_{\mathrm{error}}(\phi) = |\braket{w^\perp|\phi}|^2 = \cos^2\phi .
\end{equation}
The closer the state is to $\ket{w}$, the smaller this probability. Evaluating it at the two ends of the interval gives the error probabilities of the two extreme trajectories,
\begin{align}
P_{\text{error}}^{\min} &= \cos^2\Big[(2r+1)\alpha - r (\delta + \Delta) \Big],
\label{eq:Pmin}\\
P_{\text{error}}^{\max} &= \cos^2\Big[(2r+1)\alpha + r (\delta + \Delta) \Big],
\label{eq:Pmax}
\end{align}
corresponding to every error pushing the rotation back (under-rotation) or forward (over-rotation). Every intermediate error history lands between the two ends of the interval, so the achievable error probabilities form a band. Two cautions apply when reading these bounds. First, $\cos^2\phi$ is not monotonic, so when the interval straddles $\pi/2$ the band reaches down to zero even though neither end does, and the labels min and max then refer to the ends of the interval rather than to the smallest and largest error. Second, when $r_0$ overshoots $\pi/2$, as it does for small registers, a little under-rotation improves the result, and $P^{\min}_{\mathrm{error}}$ first decreases with the noise before it increases.

At the noiseless optimum, where $(2r_0+1)\alpha\approx\pi/2$, Eqs.~\eqref{eq:Pmin} and \eqref{eq:Pmax} give the same worst-case value,
\begin{equation}
P_{\mathrm{error}}\big|_{r_0} \approx \sin^2\!\big[r_0(\delta+\Delta)\big] \approx \Big[\frac{\pi}{4}\sqrt{N}\,(\delta+\Delta)\Big]^2 ,
\label{eq:Perr_r0}
\end{equation}
the last form holding when the argument is small. The worst-case loss is quadratic in $\sqrt{N}(\delta+\Delta)$. This is milder than the depolarizing case, where the loss was linear in $p\sqrt{N}$ in the exponent, but the scaling with the database size is the same: to keep the worst-case error fixed, the angular error per iteration must shrink as $1/\sqrt{N}$. This is the scaling found by Long and co-workers for systematic errors in the phase inversions \cite{long2000}. For errors with random rather than systematic signs the angular spread grows as $\sqrt{r}$ instead of $r$, so after $r_0\sim\sqrt{N}$ iterations it is of order $N^{1/4}(\delta+\Delta)$, and the tolerable error per iteration falls only as $N^{-1/4}$. This is the scaling found by Shenvi, Brown, and Whaley for random phase errors in the oracle \cite{shenvi2003}. A random-matrix analysis of static systematic noise on all qubits gives a threshold of the same order as the systematic case, up to a logarithm \cite{dowarah2025}. The bounds derived here are the guaranteed ones: they hold for every sequence of errors within the given limits.

\subsection{Results and discussion}

\begin{figure}[t]
    \centering
    \includegraphics[width=\linewidth]{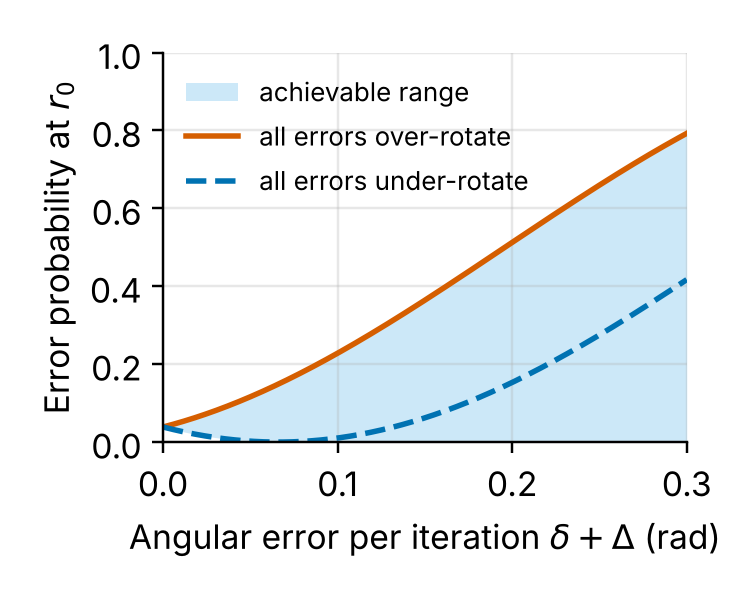}
    \caption{Error probability after $r_0=3$ iterations for a register of $n=4$ qubits ($N=16$), as a function of the total angular error per iteration, $\delta+\Delta$. The solid and dashed curves are the two ends of the angular interval, Eqs.~\eqref{eq:Pmax} and \eqref{eq:Pmin}: every error over-rotates, or every error under-rotates. The shaded region is the full range of error probabilities that bounded angular errors can produce. The under-rotated end reaches zero at $\delta+\Delta\approx0.066$~rad, where under-rotation exactly cancels the small overshoot of $r_0$. Beyond that point the interval contains $\pi/2$, and the range extends down to zero. Plotting code written with AI assistance (see Acknowledgments).}
    \label{fig:e vs n}
\end{figure}

\begin{figure}[t]
    \centering
    \includegraphics[width=\linewidth]{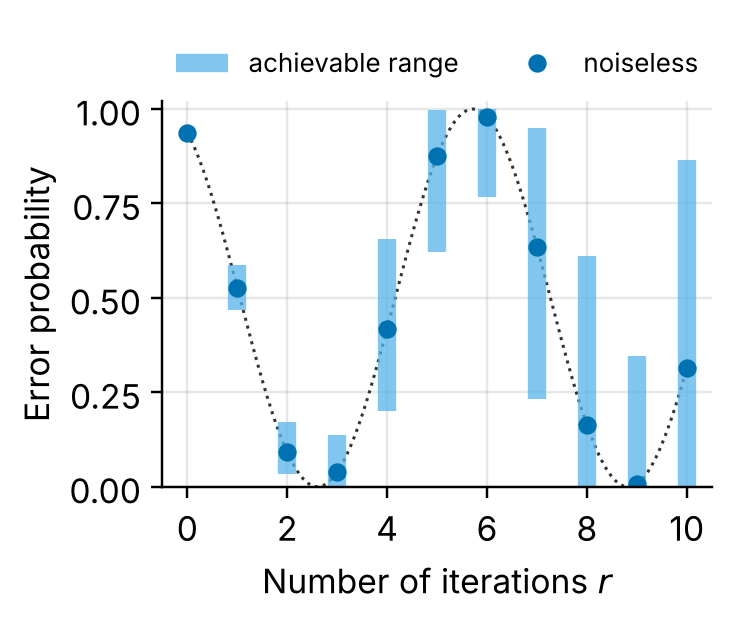}
    \caption{Error probability as a function of the number of iterations $r$ for a register of $n=4$ qubits ($N=16$) at fixed angular error, $\delta+\Delta=0.06$~rad per iteration. Dots: the noiseless case. The dotted line, $\cos^2[(2r+1)\alpha]$, oscillates with period $\pi/(2\alpha)$ in $r$. Bars: the full range of error probabilities allowed by Eqs.~\eqref{eq:Pmin} and \eqref{eq:Pmax} at each $r$. The angular interval behind each bar widens linearly with $r$, Eq.~\eqref{eq:width}, so later minima are less and less reliable. Plotting code written with AI assistance (see Acknowledgments).}
    \label{fig:e vs r}
\end{figure}

Figures~\ref{fig:e vs n} and \ref{fig:e vs r} show the error probabilities allowed by Eqs.~\eqref{eq:Pmin} and \eqref{eq:Pmax} for a small register, for which the integer $r_0$ slightly overshoots the ideal angle. The first figure varies the noise at fixed depth; the second varies the depth at fixed noise. Both use a register of $n=4$ qubits ($N=16$). Then $\alpha=\arcsin(1/4)\approx0.253$~rad and $r_0=3$, so $(2r_0+1)\alpha$ exceeds $\pi/2$ by about $0.2$~rad, and the noiseless error probability at $r_0$ is about $0.04$. The bounds depend on $\delta$ and $\Delta$ only through their sum, so we plot against $\delta+\Delta$. The shaded range in each figure is the exact set of values of $\cos^2\phi$ for $\phi$ in $[\theta^{r}_{\min},\theta^{r}_{\max}]$. It reaches zero whenever this interval contains $\pi/2$.

Figure~\ref{fig:e vs n} shows how the range opens up as the angular error grows. At zero noise the two ends coincide at the noiseless error, which is small but not zero because $r_0$ overshoots slightly for this register. Over-rotation makes things steadily worse, so the upper curve rises from the start. Under-rotation at first cancels the overshoot. The under-rotated end reaches zero error at $\delta+\Delta\approx0.066$~rad and then rises again. From that point on, the angular interval contains $\pi/2$. Some error history then still lands on the marked state, so the lower edge of the range stays at zero while the upper edge keeps rising. The width of the range, not just its center, is the message of the figure: even when the average performance is acceptable, a single run can land anywhere inside it.

Figure~\ref{fig:e vs r} shows the complementary dependence on depth. In the noiseless case the error probability oscillates and nearly vanishes at the optimal iteration counts. With bounded angular noise the range is narrow at small $r$, and the angular interval behind it widens in proportion to $r$, Eq.~\eqref{eq:width}. The first minimum, at $r=3$, is still well defined: the error stays below about $0.14$. At the second minimum, $r=9$, it can lie anywhere between zero and about $0.35$, and a run stopped there may do far worse than the noiseless curve suggests. This is the coherent counterpart of the damping in Sec.~\ref{sec:global}: there the signal shrank with depth, here it becomes uncertain with depth. In both cases the lesson is to stop at the first optimum and to keep the circuit as shallow as the database allows.

Taken together, the two figures show the trade-off that coherent errors impose. More iterations bring the ideal state closer to the marked item, but every iteration adds to the angular uncertainty. The quadratic speedup therefore becomes fragile as either the noise strength or the circuit depth grows, and the tolerable angular error per iteration must decrease as $1/\sqrt{N}$ if the search is to remain reliable for larger databases.

\section{\label{sec:concl}Conclusions}

We have followed Grover's algorithm through local incoherent noise and through coherent rotation errors, with the noise placed after every gate layer of the same circuit.

For a depolarizing channel that acts on the whole register, the channels commute with the gates and collapse into a single weight that scales the noiseless state. The success probability is then a damped Grover oscillation on top of the random-guess floor. The rotation is not slowed; its weight decays. As a result the best stopping point moves earlier as the noise grows, and the expected number of oracle queries keeps its square-root scaling only while the noise accumulated over the whole circuit stays small. Beyond that point the cost grows linearly with the database size, as the known lower bounds require, and the quadratic speedup is lost.

For independent errors on the qubits, the key observation is that the Grover operator is a simple sign flip outside the plane in which the search takes place. Amplitude that an error pushes out of the plane is frozen and lost to the search. A phase flip keeps the marked component and discards the rest, so it does most harm early in the search. A bit flip keeps the unmarked component and discards the marked one, so it does most harm late. A combined error discards both. For a large database the same observation turns the noisy search into Bloch-type equations for the part of the state that stays in the plane, with the algorithm as a rotation and the noise as leakage and dephasing at rates that follow from counting which errors remove which component. Solving them shows that local noise acts, to leading order, like a global channel strengthened by the number of qubits, and that it dephases the plane faster than it drains it. Amplitude damping is the gentlest of the channels considered, and its effect depends on the label of the marked item. In every case the tolerable error per qubit and gate layer falls as $1/(n\sqrt{N})$: one factor from the depth of the circuit and one from the number of qubits that can fail.

For coherent rotation errors of bounded size the state stays pure but its angle becomes uncertain. The uncertainty grows linearly with the number of iterations, and the success probability lies inside a band whose edges we give in closed form. Such errors do not only lower the average performance; they make individual runs unpredictable, and increasingly so with depth.

The three mechanisms act differently, by shrinking the signal, by freezing part of it, or by blurring it, but they lead to the same practical rule: the error accumulated over the whole circuit must stay of order one for the search to keep its advantage. The expressions derived here are simple enough to serve as a first estimate of the noise budget of a Grover search of a given size, and they provide a baseline against which error mitigation for amplitude amplification can be judged. The same approach can be extended to correlated or time-dependent errors and to other algorithms built on repeated amplitude amplification.

\section*{Data availability}

This is a purely theoretical work, and no data were created or analyzed in this study. All figures can be reproduced directly from the equations presented in the paper.

\begin{acknowledgments}
During this work the authors used Claude (Anthropic), an AI assistant. It was used to revise and edit the text; to check the original derivations and to develop and extend the analytical results. The authors directed this work and chose the questions and models studied.  The authors take full responsibility for the content of this paper.
\end{acknowledgments}

\bibliography{references}

\end{document}